\documentclass[english,aps,manuscript]{revtex4}
\usepackage[T1]{fontenc}
\usepackage[latin9]{inputenc}
\usepackage[letterpaper]{geometry}
\usepackage{color}
\usepackage{textcomp}
\usepackage{amsthm}
\usepackage{amsmath}
\usepackage{esint}

\makeatletter
\@ifundefined{textcolor}{}
{%
 \definecolor{BLACK}{gray}{0}
 \definecolor{WHITE}{gray}{1}
 \definecolor{RED}{rgb}{1,0,0}
 \definecolor{GREEN}{rgb}{0,1,0}
 \definecolor{BLUE}{rgb}{0,0,1}
 \definecolor{CYAN}{cmyk}{1,0,0,0}
 \definecolor{MAGENTA}{cmyk}{0,1,0,0}
 \definecolor{YELLOW}{cmyk}{0,0,1,0}
 }
\numberwithin{equation}{section}
\numberwithin{figure}{section}

\makeatother

\usepackage{babel}

\begin{document}

\title{{\huge On the validity of the Onsager relations in relativistic binary
mixtures}}

\author{Valdemar Moratto$^{1}$, A. L. Garcia-Perciante$^{2}$, L. S. Garcia-Colin$^{1\,\text{and}\,3}$}

\address{$^{1}$Depto. de Fisica, Universidad Autonoma Metropolitana-Iztapalapa,
Av. Purisima y Michoacan S/N, Mexico D. F. 09340, Mexico.}

\address{$^{2}$Depto. de Matematicas Aplicadas y Sistemas, Universidad Autonoma
Metropolitana-Cuajimalpa, Artificios 40 Mexico D.F 01120, Mexico.}

\address{$^{3}$El Colegio Nacional, Luis Gonzalez Obregon 23, Centro Historico,
Mexico D. F. 06020, Mexico.\\
 }
\begin{abstract}
In this work we study the properties of a relativistic mixture of
two non-reacting dilute species in thermal local equilibrium. Following
the conventional ideas in kinetic theory, we use the concept of chaotic
velocity. In particular, we address the nature of the density, or
pressure gradient term that arises in the solution of the linearized
Boltzmann equation in this context. Such effect, also present for
the single component problem, has so far not been analyzed from the
point of view of the Onsager resciprocity relations. In order to address
this matter, we propose two alternatives for the Onsagerian matrix
which comply with the corresponding reciprocity relations and also
show that, as in the non-relativistic case, the chemical potential
is not an adequate thermodynamic force. The implications of both representations
are briefly analyzed. 

PACS: 05.20.Dd, 03.30.+p, 05.70.Ln.
\end{abstract}
\maketitle

\section{Introduction}

Relativistic kinetic theory has become a rather fashionable subject
in recent years. Not only this is due to many astrophysical and cosmological
phenomena which occur in dilute gases at high temperatures, buy also
because for a while, it was believed it would find an important application
in the study of the quark gluon plasma which is formed in relativistic
heavy ion collisions (RICH). Although this last feature is questionable
\cite{Thoma,Schafer,Aad} mostly because a relativistic quantum hydrodynamical
theory is required, interest still remains due to its applications
to classical systems. 

It is our view that in spite of the existence of the wealth of approaches
to this classical problem, which goes back to Israel et. al. \cite{ISRAEL,Israel-Stewart,Groot-Leewen-Weert,CERCIGNANI},
there are two aspects that have been so far ignored in previous works.
First, the formulation of the theory using the rather useful concept
of chaotic (thermal) velocities of the molecules composing the gas.
Secondly, the study of the so called cross effects in irreversible
thermodynamics. Also, and even more important, the possibility of
selecting appropriate representations of fluxes and their conjugate
forces in which one can provide an airtight-proof on the validity
on Onsager's reciprocity relations (ORR) has to our knowledge, never
been given. It is important to emphasize at this point that the validity
of the ORR is one of the fundamental postulates of Linear Irreversible
Thermodynamics (LIT).

The introduction of the concept of thermal velocity has been successfully
accomplished for a single component dilute gas and its advantages
clearly underlined in the calculation of their transport properties
\cite{Sandoval - Garcia-Colin 2000,Ana}. Perhaps it is worth stressing
that in this formulation the relativistic generalization of the classical
expression for the heat flux obtained emphasizes the nature of heat
namely, the transport of the kinetic or thermal energy of the molecules.
Further, one can obtain in a rigorous way the expression for the relativistic
stress tensor as proposed phenomenologically by C. Eckart in 1940
\cite{Eckart 1 2 3}.

In this paper we study the second feature as mentioned above, the
cross effects and the validity of the ORR in a binary non-reactive
diluted mixture of gases. The most surprising result is that there
are two representations in which the ORR hold true, depending on how
fluxes and forces are selected. One of the representations follows
the idea formulated by previous authors of coupling in one single
force both the temperature and pressure gradients, this force being
the direct drive for the heat flux. The ORR are verified in that context.
The second one is based on the novel idea that due to the non-invariance
of the volume elements of the gas under Lorentz transformations, a
{}``volume flux'' results whose conjugate force is the pressure
gradient. Resemblance to this idea arose in at least one phenomenological
derivation of Burnett's constitutive equations and has also been subject
of reformulations of classical hydrodynamics \cite{Brenner,Brenner-1,Brenner-2}.
Further, two new cross effects are present in this approach which
are completely absent in the non-relativistic case. 

To accomplish this task, we divide the article as follows: Section
2 is devoted to the basic concepts of the relativistic kinetic theory,
as well as the derivation of the conservation equations. In section
3 we use the Chapman-Enskog method to linearize the Boltzmann equation.
In section 4 we select the appropriate thermodynamic forces following
the ideas in Refs. \cite{Groot-Leewen-Weert,CERCIGNANI}, and we show
that the Onsager reciprocity relations \cite{Onsager,Onsager-Machlup,CASIMIR}
in a $2\times2$ matrix hold. In section 5 we propose the new idea
of a new purely relativistic flux directly coupled with the pressure
gradient, which satisfies the symmetry of a $3\times3$ {}``Onsagerian''
matrix and further, we show that the chemical potential in this context
is not an appropriate thermodynamic force. Finally, in section 6,
we include a discussion and concluding remarks.

\section{Relativistic Kinetic Theory}

As mentioned above, we study a relativistic, dilute mixture of two
non-reacting species in thermal local equilibrium. In the framework
of kinetic theory, we consider the quantity\begin{equation}
f_{(1)}d^{3}xd^{3}v_{(1)}+f_{(2)}d^{3}xd^{3}v_{(2)}\label{eq:1}\end{equation}
which represents the number of particles of species $(1)$ and $(2)$
in $d^{3}xd^{3}v_{(1)}$ and $d^{3}xd^{3}v_{(2)}$, where $v_{(i)}^{\alpha}$
denotes molecular velocity. To establish a clear notation, we use
parenthesis in the subscripts to denote species. For components, Latin
subscripts run form 1 to 3 for the spatial ones while Greek subscripts
are used for four-vectors and tensors running from 1 to 4 in Minkowski's
space-time with a +++- signature. 

The invariant Boltzmann equations for the mixture are,\begin{equation}
v_{(i)}^{\alpha}f_{(i),\alpha}=\sum_{i,j=1}^{2}J_{(ij)}\label{eq:2}\end{equation}
where the collisional term is given by \cite{CERCIGNANI},\begin{equation}
\sum_{i,j=1}^{2}J_{(ij)}=\sum_{i,j=1}^{2}\int\left(f'_{(i)}f'_{(j)}-f_{(i)}f_{(j)}\right)F_{(ij)}\sigma_{(ij)}d\Omega_{(ji)}d^{3}v_{(j)}^{*}.\label{eq:3}\end{equation}
Here, $F_{(ij)}$, $\sigma_{(ij)}$ and $d\Omega_{(ji)}$ denote the
invariant flux, the invariant differential elastic cross-section and
the element of solid angle that characterize a binary collision between
the particles of constituent $i$ with those of constituent $j$,
respectively. The differential $d^{3}v_{(i)}^{*}$ stands for $\frac{d^{3}v_{(i)}}{v_{(i)}^{4}}$
, also an invariant. The cross-section $\sigma_{(ij)}$ has special
symmetries \cite{Teo H Weert-Leeuwen-Groot} that guarantee the existence
of inverse collisions such that the principle of microscopic reversibility
is satisfied. The quantities $f{}_{(i)}$ and $f'_{(i)}$ denote the
distribution functions before and after a collision, respectively. 

The collisional invariants in this case are the rest mass of each
species $m_{(i)}$ and the four-momentum $m_{(i)}v_{(i)}^{\alpha}$,
where the energy is included in the temporal component $m_{(i)}v_{(i)}^{4}$.
In the following subsections, these quantities will be used to obtain
balance equations. It is important to notice at this point that the
molecular velocity in the previous equations is measured by an observer
in an arbitrary frame, which we call laboratory frame.

\subsection{Particle Number Conservation }

By multiplying the Boltzmann equation, Eq. (\ref{eq:2}), by $m_{(i)}$
and integrating over $d^{3}v_{(i)}^{*}$ one finds\begin{equation}
\left(m_{(i)}\int v_{(i)}^{\alpha}f_{(i)}d^{3}v_{(i)}^{*}\right)_{,\alpha}=0,\label{eq:4}\end{equation}
where \begin{equation}
N_{(i)}^{\alpha}=m_{(i)}\int v_{(i)}^{\alpha}f_{(i)}d^{3}v_{(i)}^{*}\label{eq:5}\end{equation}
is the mass four-flux in an arbitrary frame. The barycentric velocity
is thus defined as\begin{equation}
nU^{\alpha}=\frac{N_{(1)}^{\alpha}}{m_{(1)}}+\frac{N_{(2)}^{\alpha}}{m_{(2)}},\label{eq:6}\end{equation}
which is consistent with Eckart's definition for the hydrodynamic
four-velocity. Here $n=n_{(1)}+n_{(2)}$ is the particle number density
and represents an invariant. We also define the relativistic diffusive
four-flux in the co-moving frame as,\begin{equation}
J_{(i)}^{\alpha}=m_{(i)}\int K_{(i)}^{\alpha}f_{(i)}d^{3}K_{(i)}^{*},\label{eq:8}\end{equation}
where $K_{(i)}^{\alpha}$ is the four-velocity of the particles of
the species $i$ measured in the co-moving frame, i.e. $K_{(i)}^{\alpha}$
is the chaotic or thermal velocity \cite{Maxwell,Clausius,Brush}.
Then $N_{(i)}^{\alpha}$ and $J_{(i)}^{\beta}$ are related by a Lorentz
transformation as follows, \begin{equation}
N_{(i)}^{\alpha}=\mathcal{L}_{\beta}^{\alpha}J_{(i)}^{\beta},\label{eq:7}\end{equation}
where $\mathcal{L}_{\beta}^{\alpha}$ is the transformation from the
co-moving frame, where $U^{m}=0,$ to an arbitrary one moving with
a four-velocity $U^{\alpha}$. 

With the help of these equations, one can find the complete particle
number conservation equation, see Ref. \cite{Val-1}. In this work
we only need them at Euler's level, because we will use the Chapman
and Enskog method up to first order in the gradients. Thus, we have
that,\begin{equation}
n_{(i)}U_{,\alpha}^{\alpha}+U^{\alpha}n_{(i),\alpha}=0.\label{eq:9}\end{equation}
for the particle number conservation.

\subsection{Momentum and Energy Balance}

In order to obtain the energy-momentum balance for the mixture, Boltzmann's
equation is now multiplied by $m_{(i)}v_{(i)}^{\alpha}$ and integrated
over $d^{3}v_{(i)}^{*}$, which yields\begin{equation}
T_{,\alpha}^{\beta\alpha}=\left(T_{(1)}^{\beta\alpha}+T_{(2)}^{\beta\alpha}\right)_{,\alpha}=0,\label{eq:10}\end{equation}
where\begin{equation}
T^{\beta\alpha}=\sum_{i}m_{(i)}\int v_{(i)}^{\beta}v_{(i)}^{\alpha}f_{(i)}d^{3}v_{(i)}^{*}.\label{eq:11}\end{equation}
In order to establish the form of the tensor $T^{\beta\alpha}$ we
recognize that, as defined in Eq. (\ref{eq:11}), it is referred to
an arbitrary reference frame. Thus, we can express it in terms of
$\tilde{T}^{\gamma\phi}$, measured in the co-moving frame defined
above, as\begin{equation}
T^{\beta\alpha}=\mathcal{L}_{\gamma}^{\beta}\mathcal{L}_{\phi}^{\alpha}\tilde{T}^{\gamma\phi},\label{eq:12}\end{equation}
where again, $\mathcal{L}_{\phi}^{\alpha}$ and $\mathcal{L}_{\gamma}^{\beta}$
are the Lorentz transformations from the co-moving frame to an arbitrary
one moving with a four-velocity $U^{\alpha}$. Following S. Weinberg
\cite{WEINBERG} and using the fact that the stress-energy tensor
is symmetric (see Eq. (\ref{eq:11}) ), we assume that in the co-moving
frame it has the form,\begin{equation}
\tilde{T}^{\beta\alpha}\ddot{=}\left(\begin{array}{cccc}
p & 0 & 0 & 0\\
0 & p & 0 & 0\\
0 & 0 & p & 0\\
0 & 0 & 0 & ne\end{array}\right)+\left(\begin{array}{cccc}
0 & 0 & 0 & q^{1}\\
0 & 0 & 0 & q^{2}\\
0 & 0 & 0 & q^{3}\\
q^{1} & q^{2} & q^{3} & 0\end{array}\right)+\left(\begin{array}{cccc}
\pi^{11} & \pi^{12} & \pi^{13} & 0\\
\pi^{12} & \pi^{22} & \pi^{23} & 0\\
\pi^{13} & \pi^{23} & \pi^{33} & 0\\
0 & 0 & 0 & 0\end{array}\right).\label{eq:13}\end{equation}
In Eq. (\ref{eq:13}) we have separated the proper equilibrium quantities
namely, the hydrostatic pressure \begin{equation}
p=\frac{1}{3}\tilde{T}^{mm},\label{eq:14}\end{equation}
and the energy density per particle\begin{equation}
ne=\tilde{T}^{44}.\label{eq:14.1}\end{equation}
On the other hand, the non equilibrium quantities are\begin{equation}
q^{m}=c\tilde{T}^{4m}=c\tilde{T}^{m4}\label{eq:15}\end{equation}
for the heat flux and \begin{equation}
\Pi^{mn}=\tilde{T}^{mn},\label{eq:16}\end{equation}
for the Navier tensor. Introducing Eq. (\ref{eq:13}) in Eq.(\ref{eq:12})
yields\begin{equation}
T^{\alpha\beta}=pg^{\alpha\beta}+\frac{1}{c^{2}}\left(p+ne\right)U^{\alpha}U^{\beta}+\frac{1}{c^{2}}\left(U^{\alpha}\mathcal{L}_{\mu}^{\beta}q^{\mu}+U^{\beta}\mathcal{L}_{\mu}^{\alpha}q^{\mu}\right)+\mathcal{L}_{\mu}^{\beta}\mathcal{L}_{\nu}^{\alpha}\Pi^{\mu\nu},\label{eq:17}\end{equation}
where $g^{\alpha\beta}$ is the metric tensor. In Eq. (\ref{eq:17})
we identify the first two terms as the relativistic energy-momentum
tensor at Euler's level. The third and fourth terms represent the
non-equilibrium generalization with the heat and viscous dissipation
terms as found from kinetic theory grounds for the single fluid in
Ref \cite{Ana}. 

We now calculate the derivative in Eq. (\ref{eq:10}) using Eq. (\ref{eq:17})
and its projection with the four-velocity namely, $U_{\mu}T_{,\nu}^{\mu\nu}$.
Neglecting all the terms which contain corrections whose order is
beyond Euler's regime \cite{Val-1}, and after laborious calculations
one finds: \begin{equation}
\tilde{\rho}\dot{U}^{\beta}+h^{\beta\nu}p_{,\nu}=0,\label{eq: 18}\end{equation}
and\begin{equation}
n\dot{e}=-pU_{,\mu}^{\mu},\label{eq:20}\end{equation}

where Eqs. (\ref{eq: 18})and (\ref{eq:20}) are the momentum and
internal energy balance equations respectively. Here, \begin{equation}
\tilde{\rho}=\sum_{i}m_{(i)}n_{(i)}G\left(z_{(i)}\right)=\tilde{\rho}_{(1)}+\tilde{\rho}_{(2)},\label{eq:19}\end{equation}
and \begin{equation}
G\left(z_{(i)}\right)=\frac{\mathcal{K}_{3}\left(\frac{1}{z_{(i)}}\right)}{\mathcal{K}_{2}\left(\frac{1}{z_{(i)}}\right)},\label{eq:21}\end{equation}
with $z_{(i)}=\frac{kT}{m_{(i)}c^{2}}$ being the well-known relativistic
parameter. The dot denotes a proper time derivative and is defined
as $\overset{.}{\left(\,\right)}=U^{\mu}\left(\,\right)_{,\mu}$.

Equation (\ref{eq:20}) is related to the temperature evolution by
assuming that the internal energy density depends only on the temperature
$e=C_{v}T$. The details of the calculations above can be found in
Refs. \cite{Ana,Val-1}.

\section{Linearization of the Boltzmann Equation}

In this section we proceed to apply the well-known Chapman-Enskog
method to linearize the covariant form of Boltzmann's equation. Following
the ideas in Ref. \cite{Ana}, we will perform all calculations in
the co-moving frame such that Eq. (\ref{eq:2}) now reads, \begin{equation}
K_{(i)}^{\alpha}f_{(i),\alpha}=\sum_{j=1}^{2}J_{(ij)},\label{eq:22}\end{equation}
where $K_{(i)}^{\alpha}$ is the four-velocity measured in such frame.
As usual, we now assume that the distribution functions $f_{(i)}\left(x^{\alpha},K_{(i)}^{\alpha},t\right)$
can be taken as functionals of the locally conserved variables namely
$f_{(i)}\left(x^{\alpha},K_{(i)}^{\alpha}|n_{(i)},U^{\alpha},T\right)$,
and further, they may be expanded in power series of a inhomogeneity
parameter around the local equilibrium distribution function $f_{(i)}^{(0)}$
defined in an arbitrary frame as \cite{JUTTNER,Chacon Dagdug Morales,Cubero-Hanggi},\begin{equation}
f_{(i)}^{(0)}=\frac{n_{(i)}}{4\pi c^{3}z_{(i)}\mathcal{K}_{2}\left(\frac{1}{z_{(i)}}\right)}\text{ exp}\left(\frac{U^{\beta}v_{(i)\beta}}{z_{(i)}c^{2}}\right),\label{eq:23}\end{equation}
which in the co-moving frame reduces to,\begin{equation}
f_{(i)}^{(0)}=\frac{n_{(i)}}{4\pi c^{3}z_{(i)}\mathcal{K}_{2}\left(\frac{1}{z_{(i)}}\right)}\text{ exp}\left(-\frac{\gamma_{k_{(i)}}}{z_{(i)}}\right),\label{eq:24}\end{equation}
where $\gamma_{k_{(i)}}=\left(1-k_{(i)}^{2}/c^{2}\right)^{-1/2}$
is the usual Lorentz factor and $k_{(i)}^{2}$ is the magnitude of
the chaotic or themal velocity. Omitting unnecessary arguments, we
resort to the linear theory \cite{Chapman-Cowling} and expand Eq.
(\ref{eq:22}) as \begin{equation}
f_{(i)}=f_{(i)}^{(0)}\left(1+\phi_{(i)}\right).\label{eq:25}\end{equation}

Substitution of Eq. (\ref{eq:25}) into (\ref{eq:22}) with the help
of the functional hypothesis and the Eqs. (\ref{eq:9}), (\ref{eq: 18})
and (\ref{eq:20}), leads to\begin{equation}
K_{(i)}^{m}\left\{ -\gamma_{k_{(i)}}\frac{1}{z_{(i)}c^{2}\tilde{\rho}}p_{,m}+\left(\ln n_{(i)}\right)_{,m}+\left[1+\frac{1}{z_{(i)}}\left(\gamma_{k_{(i)}}-G\left(z_{(i)}\right)\right)\right]\left(\ln T\right)_{,m}\right\} =\left[C\left(\phi_{(i)}\right)+C\left(\phi_{(i)}+\phi_{(j)}\right)\right].\label{eq:26}\end{equation}
Notice that in Eq. (\ref{eq:26}) we have omitted the second rank
tensorial terms since Curie's principle establishes that in isotropic
systems only forces and fluxes of the same tensorial rank couple among
themselves. Clearly, there is an equation similar to Eq. (\ref{eq:26})
for species $j$. The linearized collision kernel now reads\begin{equation}
C\left(\phi_{(i)}+\phi_{(j)}\right)=\int\cdots\int f_{(i)}^{(0)}f_{(j)}^{(0)}\left(\phi_{(j)}\text{\textasciiacute}+\phi_{(i)}\text{\textasciiacute}-\phi_{(j)}-\phi_{(i)}\right)F_{(ij)}\sigma_{(ij)}d\Omega_{(ji)}d^{3}v_{(j)}^{*},\label{eq:27}\end{equation}
and, \[
C\left(\phi_{(i)}\right)=\int\cdots\int f_{(i)}^{(0)}f_{(i)}^{(0)}\left(\phi_{(i)}\text{\textasciiacute}+\phi_{(i)}\text{\textasciiacute}-\phi_{(i)}-\phi_{(i)}\right)F_{(ii)}\sigma_{(ii)}d\Omega_{(ii)}d^{3}v_{(i)}^{*}.\]

The left hand side of Eq. (\ref{eq:26}) contains terms involving
gradients of the intensive thermodynamical variables, $p_{,m}$, $\left(n_{(i)}\right)_{,m}$
and $T_{,m}$, which we identify with thermodynamic forces. The question
that arises is how to select among them, a representation in which
Onsager's reciprocity relations hopefully turn out to be valid. This
will be discussed in the following sections.

\section{Solution With Two Thermodynamic Forces }

Following the statement issued above, we will proceed to discuss the
aforementioned representations. For instance, we first re-arrange
the left hand side of Eq. (\ref{eq:26}) to read as,\begin{equation}
K_{(i)}^{m}\left\{ \left[d_{m(ij)}\right]+\frac{1}{z_{(i)}}\left(\gamma_{k_{(i)}}-G\left(z_{(i)}\right)\right)\left[\frac{T_{,m}}{T}-\frac{1}{nh_{E}}p_{,m}\right]\right\} =\left[C\left(\phi_{(i)}\right)+C\left(\phi_{(i)}+\phi_{(j)}\right)\right],\label{eq:28}\end{equation}
where \begin{equation}
d_{m(ij)}=n_{(j)}\left(\frac{m_{(j)}G\left(z_{(j)}\right)-m_{(i)}G\left(z_{(i)}\right)}{\tilde{\rho}}\right)\frac{p_{,m}}{p}+\frac{n}{n_{(i)}}\left(n_{i0}\right)_{,m},\label{eq:29}\end{equation}
and using the notation $nh_{E}=\tilde{\rho}c^{2}$, and $n_{i0}=\frac{n_{(i)}}{n}$
representing an invariant. This choice implies that we are considering
two vector forces in the system namely,\begin{equation}
d_{m(ij)}\qquad\text{and}\qquad\frac{T_{,m}}{T}-\frac{1}{nh_{E}}p_{,m}.\label{eq:30}\end{equation}
The second term may be regarded as related to a generalized Fourier's
equation with a thermal force that includes both a temperature and
pressure gradients \cite{CERCIGNANI}. Further it may be shown that
in the non-relativistic limit, the coefficient of such force in Eq.
(\ref{eq:28}) reduces to, \begin{equation}
\frac{1}{z_{(i)}}\left(\gamma_{k_{(i)}}-G\left(z_{(i)}\right)\right)\rightarrow\frac{m_{(i)}k_{(i)}^{2}}{2k_{B}T}-\frac{5}{2},\end{equation}
and the coefficient of the pressure gradient vanishes because $\left(\gamma_{k_{(i)}}-G\left(z_{(i)}\right)\right)\rightarrow0$.
Thus, the inhomogeneous term in Eq. (\ref{eq:26}) reduces to the
well-known expression of the classical linearized Boltzmann equation. 

On the other hand $d_{m(ij)}=-d_{m(ji)}\equiv d_{m}$ may be considered
as a generalization of the standard diffusive force to a relativistic
scheme, since indeed, in the non relativist case Eq. (\ref{eq:29})
reduces to,\begin{equation}
d_{m(ij)}\rightarrow\frac{n_{(j)}}{\rho p}\left(m_{(j)}-m_{(i)}\right)\nabla p+\frac{n}{n_{(i)}}\nabla n_{i0},\label{eq:31}\end{equation}
which is in accordance with phenomenological \cite{Groot Mazur} and
kinetic \cite{On the validity of the Onsager relations...} classical
expressions. 

Having selected the above thermodynamic forces, the solution to Eq.
(\ref{eq:28}) reads as \cite{Ana 2008,Curtis}, \begin{equation}
\phi_{(i)}=-K_{(i)}^{m}A_{(i)}\left[\frac{T_{,m}}{T}-\frac{1}{nh_{E}}p_{,m}\right]-\sum_{i}K_{(i)}^{m}D_{(i)}d_{m}.\label{eq:32}\end{equation}

Substitution of Eq. (\ref{eq:32}) in (\ref{eq:28}) leads to two
independent equations for the scalar functions $A_{(i)}$ and $D_{(i)}$,
namely,\begin{equation}
K_{(i)}^{m}=-\sum_{j}\int\cdots\int f_{(i)}^{(0)}f_{(j)}^{(0)}\left[K_{(j)}^{m}\text{\textasciiacute}D_{(j)}\text{\textasciiacute}+K_{(i)}^{m}\text{\textasciiacute}D_{(i)}\text{\textasciiacute}-K_{(j)}^{m}D_{(j)}-K_{(i)}^{m}D_{(i)}\right]F_{(ij)}\sigma_{(ij)}d\Omega_{(ji)}d^{3}K_{(j)}^{*},\label{eq:33}\end{equation}
\begin{equation}
\begin{array}{c}
K_{(i)}^{m}\frac{1}{z_{(i)}}\left(\gamma_{k_{(i)}}-G\left(z_{(i)}\right)\right)=\\
-\sum_{j}\int\cdots\int f_{(i)}^{(0)}f_{(j)}^{(0)}\left[K_{(j)}^{m}\text{\textasciiacute}A_{(j)}\text{\textasciiacute}+K_{(i)}^{m}\text{\textasciiacute}A_{(i)}\text{\textasciiacute}-K_{(j)}^{m}A_{(j)}-K_{(i)}^{m}A_{(i)}\right]F_{(ij)}\sigma_{(ij)}d\Omega_{(ji)}d^{3}K_{(j)}^{*},\end{array}\label{eq:34}\end{equation}
We will now use the expressions for the mass and energy fluxes arising
in this representation to prove the validity of Onsager's reciprocity
relations in this scheme. The diffusive mass flux has been defined
in Eq. (\ref{eq:8}), which with the help of Eq. (\ref{eq:32}) and
(\ref{eq:25}) can be written as follows,\begin{equation}
\frac{J_{(i)}^{m}}{m_{(i)}}=-\frac{1}{3}\int f_{(i)}^{(0)}K_{(i)}^{n}K_{n(i)}A_{(i)}d^{3}K_{(i)}^{*}\left[\frac{T^{,m}}{T}-\frac{1}{nh_{E}}p^{,m}\right]-\frac{1}{3}\int f_{(i)}^{(0)}K_{(i)}^{n}K_{n(i)}D_{(i)}d^{3}K_{(i)}^{*}d^{m},\label{eq:35}\end{equation}
In Eq. (\ref{eq:35}) the transport coefficients are identified as,\begin{equation}
\frac{J_{(i)}^{m}}{m_{(i)}}=-L_{dq}\left[\frac{T^{,m}}{T}-\frac{1}{nh_{E}}p^{,m}\right]-L_{dd}d^{m},\label{eq:36}\end{equation}
where $L_{dq}$ and $L_{dd}$ are the integrals appearing in Eq. (\ref{eq:35}).

For the energy flux we propose the form which is given in the literature
\cite{Groot Mazur}, \begin{equation}
\frac{q_{tot}^{m}}{kT}=\frac{1}{kT}\sum_{i}\left(q_{(i)}^{m}-h_{(i)}J_{(i)}^{m}\right),\label{eq:37}\end{equation}
where\begin{equation}
h_{(i)}=\frac{kT}{z_{(i)}}G\left(z_{(i)}\right),\label{eq:38}\end{equation}
is the enthalpy \cite{Groot-Leewen-Weert}. After Eqs. (\ref{eq:15})
and (\ref{eq:8}) are introduced in Eq. (\ref{eq:37}) one obtains,
\begin{eqnarray}
\frac{q_{tot}^{m}}{kT} & = & -\frac{1}{3}\sum_{i}\int f_{(i)}^{(0)}\frac{1}{z_{(i)}}\left(\gamma_{k_{(i)}}-G\left(z_{(i)}\right)\right)K_{(i)}^{n}K_{n(i)}A_{(i)}d^{3}K_{(i)}^{*}\left[\frac{T^{,m}}{T}-\frac{1}{nh_{E}}p^{,m}\right]\label{eq:39}\\
 &  & -\frac{1}{3}\sum_{i}\int f_{(i)}^{(0)}\frac{1}{z_{(i)}}\left(\gamma_{k_{(i)}}-G\left(z_{(i)}\right)\right)K_{(i)}^{n}K_{n(i)}D_{(i)}d^{3}K_{(i)}^{*}d^{m},\end{eqnarray}
or\begin{equation}
\frac{q_{tot}^{m}}{kT}=-L_{qq}\left[\frac{T^{,m}}{T}-\frac{1}{nh_{E}}p^{,m}\right]-L_{qd}d^{m}.\label{eq:40}\end{equation}
Equations (\ref{eq:36}) and (\ref{eq:40}) are now in a form which,
by a similar analysis as the one performed in the classical case (see
Ref. \cite{On the validity of the Onsager relations...}) are bound
to lead to the required relations of symmetry. 

To show this we start by constructing an {}``Onsagerian'' matrix,
namely,\begin{equation}
\left(\begin{array}{c}
q_{tot}^{m}\\
J_{(i)}^{m}\end{array}\right)=-\left(\begin{array}{cc}
L_{qq} & L_{qd}\\
L_{dq} & L_{dd}\end{array}\right)\left(\begin{array}{c}
\frac{T^{,m}}{T}-\frac{1}{nh_{E}}p^{,m}\\
d^{m}\end{array}\right).\label{eq:41}\end{equation}

Then, one proceeds by multiplying both sides of Eq. (\ref{eq:33})
by $K_{(i)m}A_{(i)}$ and integrating over $d^{3}K_{(i)}^{*}$ to
obtain the form\begin{equation}
\begin{array}{c}
\int\frac{1}{3}\left(K_{(i)}^{n}K_{n(i)}\right)A_{(i)}d^{3}K_{(i)}^{*}\\
=-\sum_{j}\int\cdots\int f_{(i)}^{(0)}f_{(j)}^{(0)}\left[K_{(j)}^{m}\text{\textasciiacute}D_{(j)}\text{\textasciiacute}+K_{(i)}^{m}\text{\textasciiacute}D_{(i)}\text{\textasciiacute}-K_{(j)}^{m}D_{(j)}-K_{(i)}^{m}D_{(i)}\right]K_{(i)m}A_{i}F_{(ij)}\sigma_{(ij)}d\Omega_{(ji)}d^{3}K_{(j)}^{*}d^{3}K_{(i)}^{*}\\
\equiv\{D,A\}.\end{array}\label{eq:42}\end{equation}
On the other hand, multiplying Eq. (\ref{eq:34}) by $K_{(i)m}D_{(i)}$
and integrating over $dK_{(i)}^{*}$ yields\begin{equation}
\begin{array}{c}
\int\frac{1}{3}\left(K_{(i)}^{n}K_{n(i)}\right)\frac{1}{z_{(i)}}\left(\gamma_{k_{(i)}}-G\left(z_{(i)}\right)\right)D_{(i)}d^{3}K_{(i)}^{*}\\
=-\sum_{j}\int\cdots\int f_{(i)}^{(0)}f_{(j)}^{(0)}\left[K_{(j)}^{m}\text{\textasciiacute}A_{(j)}\text{\textasciiacute}+K_{(i)}^{m}\text{\textasciiacute}A_{(i)}\text{\textasciiacute}-K_{(j)}^{m}A_{(j)}-K_{(i)}^{m}A_{(i)}\right]K_{(i)m}D_{(i)}F_{(ij)}\sigma_{(ij)}d\Omega_{(ji)}d^{3}K_{(j)}^{*}d^{3}K_{(i)}^{*}\\
\equiv\{A,D\}.\end{array}\label{eq:43}\end{equation}

Equations (\ref{eq:42}) and (\ref{eq:43}) may be symmetrized by
taking into account the invariance of $F_{(ij)}\sigma_{(ij)}d\Omega_{(ji)}d^{3}K_{(j)}^{*}d^{3}K_{(i)}^{*}$
(see Ref. \cite{CERCIGNANI}), and using the same symmetry arguments
as in the conventional proof of the H-theorem. Such a procedure leads
to\begin{equation}
\{A,D\}=\{D,A\},\label{eq:44}\end{equation}
and thus\begin{equation}
L_{dq}=L_{qd}.\label{eq:45}\end{equation}

Emphasis should be made on the fact that we verified the reciprocity
of the Onsager relations using the standard kinetic definition for
two fluxes but not for the forces. In this section, we are assuming
that the generalization for the Fourier's equation has the form given
by Eq. (\ref{eq:40}). Here $p^{,m}$ is considered as part of this
force in order to obtain integral equations in which the transformation
of their kernels fulfill the symmetry requirements. Thus in this representation
one cannot speak of the canonical forms of the Dufour-Soret effects
that relate the diffusion coefficients to \emph{strictly the thermal
conductivity}.

However, in the following section we will overcome this difficulty
by introducing a volumetric flow which arises solely from the fact
that in the theory of relativity volumes are not invariants. This
representation is completely new and bears some resemblance with recent
work by H. Brenner \cite{Brenner,Brenner-1,Brenner-2}, who argues
that this kind of fluxes are important in non-relativistic fluids.

\section{Solution With Three Thermodynamic Forces}

In this section we explore the possibility of a third thermodynamic
flux in the system. The motivation behind such task is the interest
to explore the possibility of keeping the temperature and pressure
gradients as independent forces which would yield a Fourier-type constitutive
equation for the heat flux relating it exclusively to a temperature
gradient. This will imply that the heat flux caused by a pressure
gradient constitutes a cross effect. This is a purely relativistic
effect and we shall see how it relates to with the pressure, or density,
gradient term that arises in the case of the high temperature in a
one component gas.

To achieve this new representation we start by re-arranging Eq. (\ref{eq:26})
as follows,\begin{equation}
K_{(i)}^{m}\left\{ d_{m}+\frac{1}{z_{(i)}}\left(\gamma_{k_{(i)}}-G\left(z_{(i)}\right)\right)\frac{T_{,m}}{T}-\left(\gamma_{k_{(i)}}-G\left(z_{(i)}\right)\right)\left[\frac{n_{(i)}m_{(i)}}{\tilde{\rho}}\frac{p_{,m}}{p_{(i)}}\right]\right\} =\left[C\left(\phi_{(i)}\right)+C\left(\phi_{(i)}+\phi_{(j)}\right)\right]\label{eq:46}\end{equation}
for species $i$, recalling that there is a similar equation for species
$j$. Notice that we are considering a new force, \[
V_{(i)m}\equiv\frac{n_{(i)}m_{(i)}}{\tilde{\rho}}\frac{p_{,m}}{p_{(i)}},\]
which satisfies \begin{equation}
V_{(1)m}=\frac{m_{(1)}}{m_{(2)}}V_{(2)m}\equiv V_{m}.\label{eq:46.1}\end{equation}
Equation (\ref{eq:46}) leads to a solution of the form\begin{equation}
\phi_{(i)}=-K_{(i)}^{m}A_{(i)}\frac{T_{,m}}{T}-\sum_{j}K_{(j)}^{m}B_{(j)}V_{m}-\sum_{j}K_{(j)}^{m}D_{(j)}d_{m}.\label{eq:47}\end{equation}

Substitution of Eq. (\ref{eq:47}) into (\ref{eq:46}) yields three
independent equations, namely,\begin{equation}
K_{(i)}^{m}\frac{1}{z_{(i)}}\left(\gamma_{k_{(i)}}-G\left(z_{(i)}\right)\right)=\left[C\left(K_{(i)}^{m}A_{(i)}\right)+C\left(K_{(i)}^{m}A_{(i)}+K_{(j)}^{m}A_{(j)}\right)\right],\label{eq:48}\end{equation}
\begin{equation}
K_{(i)}^{m}\left(\gamma_{k_{(i)}}-G\left(z_{(i)}\right)\right)=\left[C\left(K_{(i)}^{m}B_{(i)}\right)+C\left(K_{(i)}^{m}B_{(i)}+K_{(j)}^{m}B_{(j)}\right)\right],\label{eq:49}\end{equation}
\begin{equation}
K_{(i)}^{m}=\left[C\left(K_{(i)}^{m}D_{(i)}\right)+C\left(K_{(i)}^{m}D_{(i)}+K_{(j)}^{m}D_{(j)}\right)\right].\label{eq:50}\end{equation}

Equation (\ref{eq:49}) is now the new ingredient in this representation.
To understand its physical meaning we proceed as follows. Consider
the motion of an individual particle which collides with another one.
After the collision it will travel a length $\lambda$, the mean free
path, before colliding with a third one. Recall also that the mean
free time is much greater than the collision time. One can thus construct
a sphere centered in the particle (in general it can be any other
geometric figure) with volume $V=\frac{4}{3}\pi\lambda^{3}$ that,
when the speed of the particle is comparable with the speed of light,
by Lorentz's contraction, is deformed into a ellipsoid with volume
$\frac{4}{3}\pi\lambda^{3}\gamma_{k}$. Therefore, in the relativistic
case, an observer sees a change in this volume with a privileged direction
$\vec{k}$. This is the process which gives rise to {}``volume or
volumetric flow''\textcolor{black}{{} and a system with an apparently
additional state variable. In order to explore its significance we
establish the transport equation characterizing its flow. In the case
of a binary mixture by multiplying Boltzmann's equation by the microscopic
change in the volume $a\gamma_{k_{(i)}}$ where $a$ is a constant,
and integrating over the velocities $d^{3}K_{(i)}^{*}$ yields,}\begin{eqnarray}
\left(\int\gamma_{k(i)}K_{(i)}^{\alpha}f_{(i)}d^{3}K_{(i)}^{*}\right)_{,\alpha} & = & \int\gamma_{k_{(i)}}\left(J_{(ii)}+J_{(ij)}\right)d^{3}K_{(i)}^{*}\label{eq: vol flow}\\
 & = & \pi_{vol}\nonumber \end{eqnarray}
which is a balance equation for the change in the volume in the gas.
Notice that in the non-relativistic limit, the right hand side vanishes,
implying that there is no such change in volume. The physical implications
of this flux are further discussed in the final section.

In the case of mixtures, the energy flux corresponding to heat dissipation
to be considered in Onsager's formalism is constructed by subtracting
the diffusive mass flux times the enthalpy from the heat flux \cite{Groot Mazur,Kinetic Theory and Irrev Thermodyn EU}.
In a similar fashion, we define the total volume (adimensional) flux
as,\begin{equation}
J_{V}^{m}=\sum_{i}\left(\int\gamma_{k_{(i)}}K_{(i)}^{m}f_{(i)}d^{3}K_{(i)}^{*}-\frac{h_{E(i)}}{m_{(i)}c^{2}}\frac{J_{(i)}^{m}}{m_{(i)}}\right),\label{eq:51}\end{equation}
where $n_{(i)}h_{E(i)}=c^{2}\tilde{\rho}_{(i)}$. Thus, using Eqs.
(\ref{eq:47}) and (\ref{eq:25}) we have that,\begin{eqnarray}
J_{V}^{m} & = & -\frac{1}{3}\sum_{i}\int f_{(i)}^{(0)}\left(\gamma_{k_{(i)}}-G\left(z_{(i)}\right)\right)K_{(i)}^{n}K_{(i)n}A_{(i)}d^{3}K_{(i)}^{*}\frac{T^{,m}}{T}\label{eq:52}\\
 &  & -\frac{1}{3}\sum_{i}\int f_{(i)}^{(0)}\left(\gamma_{k_{i}}-G\left(z_{(i)}\right)\right)K_{(i)}^{n}K_{(i)n}B_{(i)}d^{3}K_{(i)}^{*}V^{m}\nonumber \\
 &  & -\frac{1}{3}\sum_{i}\int f_{(i)}^{(0)}\left(\gamma_{k_{(i)}}-G\left(z_{(i)}\right)\right)K_{(i)}^{n}K_{(i)n}D_{(i)}d^{3}K_{(i)}^{*}d^{m}\nonumber \end{eqnarray}
or\begin{equation}
J_{V}^{m}=-L_{Vq}\frac{T^{,m}}{T}-L_{VV}V^{m}-L_{Vd}d^{m}.\label{eq:53}\end{equation}
which introduces two new transport cross-coefficients $L_{Vq}$, $L_{Vd}$
and one corresponding to the direct effect $L_{VV}$. 

As mentioned before, the dissipative energy flux is given by \begin{eqnarray}
\frac{q_{tot}^{m}}{k_{B}T} & = & -\frac{1}{3}\sum_{i}\int f_{(i)}^{(0)}\frac{1}{z_{(i)}}\left(\gamma_{k_{(i)}}-G\left(z_{(i)}\right)\right)K_{(i)}^{n}K_{n(i)}A_{(i)}d^{3}K_{(i)}^{*}\frac{T^{,m}}{T}\\
 &  & -\frac{1}{3}\sum_{i}\int f_{(i)}^{(0)}\frac{1}{z_{(i)}}\left(\gamma_{k_{(i)}}-G\left(z_{(i)}\right)\right)K_{(i)}^{n}K_{n(i)}B_{(i)}d^{3}K_{(i)}^{*}V^{m}\nonumber \\
 &  & -\frac{1}{3}\sum_{i}\int f_{(i)}^{(0)}\frac{1}{z_{(i)}}\left(\gamma_{k_{(i)}}-G\left(z_{(i)}\right)\right)K_{(i)}^{n}K_{n(i)}D_{(i)}d^{3}K_{(i)}^{*}d^{m}\nonumber \end{eqnarray}
or\begin{equation}
\frac{q_{tot}^{m}}{k_{B}T}=-L_{qq}\frac{T^{,m}}{T}-L_{qV}V^{m}-L_{qd}d^{m}.\label{eq:53.1}\end{equation}

and for the mass flow we have,\begin{equation}
\frac{J_{(i)}^{m}}{m_{(i)}}=-\frac{1}{3}\int f_{(i)}^{(0)}K_{(i)}^{n}K_{n(i)}A_{(i)}d^{3}K_{(i)}^{*}\frac{T^{,m}}{T}-\frac{1}{3}\int f_{(i)}^{(0)}K_{(i)}^{n}K_{n(i)}B_{(i)}d^{3}K_{(i)}^{*}V^{m}-\frac{1}{3}\int f_{(i)}^{(0)}K_{(i)}^{n}K_{n(i)}D_{(i)}d^{3}K_{(i)}^{*}d^{m},\end{equation}
which can also be written as\begin{equation}
\frac{J_{(i)}^{m}}{m_{(i)}}=-L_{dq(i)}\frac{T^{,m}}{T}-L_{dV(i)}V^{m}-L_{dd(i)}d^{m}.\label{eq:53.2}\end{equation}
From the previous equations, one can readily identify the Soret and
Dufour cross-effects. The verification of the Onsager reciprocity
relations will support that these are the correct generalizations
for such effects. 

Equations (\ref{eq:53}), (\ref{eq:53.1}) and (\ref{eq:53.2}) will
be explored to see whether they comply with the Onsager reciprocity
relations. As before, we construct the Onsagerian matrix,\begin{equation}
\left(\begin{array}{c}
q_{tot}^{m}\\
J_{(i)}^{m}\\
J_{V}^{m}\end{array}\right)=-\left(\begin{array}{ccc}
L_{qq} & L_{qd} & L_{qV}\\
L_{dq(i)} & L_{dd(i)} & L_{dV(i)}\\
L_{Vq} & L_{Vd} & L_{VV}\end{array}\right)\left(\begin{array}{c}
\frac{T^{,m}}{T}\\
d_{(i)}^{m}\\
V^{m}\end{array}\right),\label{eq:54}\end{equation}
where we introduced the term $V^{m}$ as the direct driving force
for the volume flux $J_{V}^{m}$. Then, by the same procedure and
arguments as those in the previous section, we will verify the symmetries\begin{equation}
L_{dq(i)}\overset{?}{=}L_{qd}\label{eq:54.1}\end{equation}
\begin{equation}
L_{Vq}\overset{?}{=}L_{qV}\label{eq:54.2}\end{equation}
\begin{equation}
L_{Vd}\overset{?}{=}L_{dV(i)}.\label{eq:54.3}\end{equation}
First, for Eq. (\ref{eq:54.1}), Eqs. (\ref{eq:48}) and (\ref{eq:50})
are multiplied by $K_{(i)}^{m}D_{(i)}$ and $K_{(i)}^{m}A_{(i)}$
respectively. After integration over $d^{3}K_{(i)}^{*}$ one finds,\begin{equation}
\begin{array}{c}
\int K_{(i)}^{n}\frac{1}{z_{(i)}}\left(\gamma_{k_{(i)}}-G\left(z_{(i)}\right)\right)K_{n(i)}D_{(i)}dK_{(i)}^{*}\\
=-\sum_{j}\int\cdots\int f_{(i)}^{(0)}f_{(j)}^{(0)}\left[K_{(j)}^{m}\text{\textasciiacute}A_{(j)}\text{\textasciiacute}+K_{(i)}^{m}\text{\textasciiacute}A_{(i)}\text{\textasciiacute}-K_{(j)}^{m}A_{(j)}-K_{(i)}^{m}A_{(i)}\right]K_{(i)}^{m}D_{(i)}F_{(ij)}\sigma_{(ij)}d\Omega_{(ji)}d^{3}K_{(j)}^{*}d^{3}K_{(i)}^{*}\\
\equiv\{A,D\},\end{array}\label{eq:54.4.1}\end{equation}
or\begin{equation}
\begin{array}{c}
\int K_{(i)}^{n}K_{n(i)}A_{(i)}dK_{(i)}^{*}\\
=-\sum_{j}\int\cdots\int f_{(i)}^{(0)}f_{(j)}^{(0)}\left[K_{(j)}^{m}\text{\textasciiacute}D_{(j)}\text{\textasciiacute}+K_{(i)}^{m}\text{\textasciiacute}D_{(i)}\text{\textasciiacute}-K_{(j)}^{m}D_{(j)}-K_{(i)}^{m}D_{(i)}\right]K_{(i)m}A_{(i)}F_{(ij)}\sigma_{(ij)}d\Omega_{(ji)}d^{3}K_{(j)}^{*}d^{3}K_{(i)}^{*}\\
\equiv\{D,A\}\end{array}\end{equation}
where, by the symmetry properties of the collisional term, $\{A,D\}=\{D,A\}$,
implying that Eq. (\ref{eq:54.1}) holds. Secondly, multiplying Eqs.
(\ref{eq:48}) and (\ref{eq:49}) by $K_{(i)}^{m}B_{(i)}$ and $K_{(i)}^{m}A_{(i)}$
respectively, and integrating over $d^{3}K_{(i)}^{*}$ yields,\begin{equation}
\begin{array}{c}
\int K_{(i)}^{n}\frac{1}{z_{(i)}}\left(\gamma_{k_{(i)}}-G\left(z_{(i)}\right)\right)K_{n(i)}B_{(i)}dK_{(i)}^{*}\\
=-\sum_{j}\int\cdots\int f_{(i)}^{(0)}f_{(j)}^{(0)}\left[K_{(j)}^{m}\text{\textasciiacute}A_{(j)}\text{\textasciiacute}+K_{(i)}^{m}\text{\textasciiacute}A_{(i)}\text{\textasciiacute}-K_{(j)}^{m}A_{(j)}-K_{(i)}^{m}A_{(i)}\right]K_{m(i)}B_{(i)}F_{(ij)}\sigma_{(ij)}d\Omega_{(ji)}d^{3}K_{(j)}^{*}d^{3}K_{(i)}^{*},\\
\equiv\{A,B\}\end{array}\end{equation}
or\begin{equation}
\begin{array}{c}
\int K_{(i)}^{n}\left(\gamma_{k_{(i)}}-G\left(z_{(i)}\right)\right)K_{n(i)}A_{(i)}dK_{(i)}^{*}\\
=-\sum_{j}\int\cdots\int f_{(i)}^{(0)}f_{(j)}^{(0)}\left[K_{(j)}^{m}\text{\textasciiacute}B_{(j)}\text{\textasciiacute}+K_{(i)}^{m}\text{\textasciiacute}B_{(i)}\text{\textasciiacute}-K_{(j)}^{m}B_{(j)}-K_{(i)}^{m}B_{(i)}\right]K_{m(i)}A_{(i)}F_{(ij)}\sigma_{(ij)}d\Omega_{(ji)}d^{3}K_{(j)}^{*}d^{3}K_{(i)}^{*},\\
\equiv\{B,A\}\end{array}\end{equation}
and since $\{A,D\}=\{B,A\}$, Eq. (\ref{eq:54.2}) holds. Lastly,
Eqs. (\ref{eq:49}) and (\ref{eq:50}) are multiplied by $K_{(i)}^{m}D_{(i)}$
and $K_{(i)}^{m}B_{(i)}$ respectively, yielding,\begin{equation}
\begin{array}{c}
\int K_{(i)}^{n}\left(\gamma_{k_{(i)}}-G\left(z_{(i)}\right)\right)K_{n(i)}D_{(i)}d^{3}K_{(i)}^{*}\\
=-\sum_{j}\int\cdots\int f_{(i)}^{(0)}f_{(j)}^{(0)}\left[K_{(j)}^{m}\text{\textasciiacute}B_{(j)}\text{\textasciiacute}+K_{(i)}^{m}\text{\textasciiacute}B_{(i)}\text{\textasciiacute}-K_{(j)}^{m}B_{(j)}-K_{(i)}^{m}B_{(i)}\right]K_{m(i)}D_{(i)}d^{3}K_{(i)}^{*}\\
\equiv\{B,D\}\end{array}\end{equation}
\begin{equation}
\begin{array}{c}
\int K_{(i)}^{n}K_{n(i)}B_{(i)}dK_{(i)}^{*}\\
=-\sum_{j}\int\cdots\int f_{(i)}^{(0)}f_{(j)}^{(0)}\left[K_{(j)}^{m}\text{\textasciiacute}D_{(j)}\text{\textasciiacute}+K_{(i)}^{m}\text{\textasciiacute}D_{(i)}\text{\textasciiacute}-K_{(j)}^{m}D_{(j)}-K_{(i)}^{m}D_{(i)}\right]K_{(i)m}B_{(i)}F_{(ij)}\sigma_{(ij)}d\Omega_{(ji)}d^{3}K_{(j)}^{*}d^{3}K_{(i)}^{*},\\
\equiv\{D,B\}\end{array}\end{equation}
where again, $\{B,D\}=\{D,B\}$, justifying Eq. (\ref{eq:54.3}) .

At this point we have verified that Onsager's symmetries hold in this
representation. The authentic Dufour effect corresponds to the transport
coefficient $L_{qd}$, while the Soret effect is related to $L_{dq}$,
whose explicit expressions are depicted in Eqs. (\ref{eq:53.1}) and
(\ref{eq:53.2}). Now, from the Onsagerian matrix we identify two
new cross-effects represented by $L_{dV}$ and $L_{qV}$. These effects
do not appear in the non-relativistic theory. Curiously enough, they
have been proposed in an entirely phenomenological way by several
authors in a non-relativistic version of linear irreversible thermodynamics
whose origin dates back to the basis of hydrodynamics as formulated
by L. Euler in 1755 Ref. \cite{Euler 1755}. However, the volume flow
is defined here by taking into account the non-invariance of an element
volume under Lorentz transformations, and vanishes in the non-relativistic
limit. In the classical framework mentioned above, its origin is entirely
different \cite{Brenner-1}.

The Onsager reciprocity relations are not necessarily fulfilled in
other representations. We shall present an example namely, the most
common case of the relativistic binary mixture where instead of using
$d^{m}$ as the thermodynamic force, the gradient of the chemical
potential is used. In classical irreversible thermodynamics one often
finds that many writers believe that the appropriate thermodynamic
forces to describe cross-effects in the case of mixtures are the chemical
potentials of the species. For the non-relativistic case, when the
mixture is non-isothermal it was clearly shown Ref. \cite{On the validity of the Onsager relations...}
that this is incorrect. In such representation the ORR do not hold
true. Here we wish to show that the same statement is valid for a
non-isothermal binary mixture of inert gases in special relativity. 

To verify this statement we recall that for an ideal gas the chemical
potential reads as,\begin{equation}
\mu_{(i)}=\frac{k_{B}T}{m_{(i)}}\left(\ln n_{(i)}-\frac{3}{2}\ln\alpha(T)\right),\end{equation}
where $\alpha(T)$ is a function related to the Jüttner distribution
which is irrelevant to the present calculation. Whence, \begin{equation}
\left(\nabla\mu_{(i)}\right)_{T}=z_{(i)}c\text{\texttwosuperior}\frac{\nabla n_{(i)}}{n_{(i)}},\label{eq:56}\end{equation}
where $\left(\nabla\mu_{(i)}\right)_{T}$ is the gradient of the chemical
potential for species$i$ at constant temperature. Further Eq. (\ref{eq:29})
can be rewritten by using the equation of state $p=nk_{B}T$ as\begin{equation}
d^{m}=\frac{n_{(j)}}{\tilde{\rho}}\left(m_{(j)}G\left(z_{(j)}\right)-m_{(i)}G\left(z_{(i)}\right)\right)\frac{T^{,m}}{T}+\frac{n_{(j)}}{\tilde{\rho}}\left(m_{(j)}G\left(z_{(j)}\right)\frac{n_{(i)}^{,m}}{n_{(i)}}-m_{(i)}G\left(z_{(i)}\right)\frac{n_{(j)}^{,m}}{n_{(j)}}\right).\label{eq:57}\end{equation}
Substitution of Eq. (\ref{eq:56}) into (\ref{eq:57}), introducing
the resulting expression for $d^{m}$ in Eqs. (\ref{eq:53}), (\ref{eq:53.1})
and (\ref{eq:53.2}) leads to,\begin{eqnarray}
\frac{q_{tot}^{,m}}{k_{B}T} & = & -\frac{1}{3}\sum_{i}\int\frac{f_{(i)}^{(0)}\left(\gamma_{k_{(i)}}-G\left(z_{(i)}\right)\right)}{z_{(i)}}K_{(i)}^{n}K_{n(i)}\left(A_{(i)}+D_{(i)}\frac{n_{(j)}}{\tilde{\rho}}\left(m_{(j)}G\left(z_{(j)}\right)-m_{(i)}G\left(z_{(i)}\right)\right)\right)d^{3}K_{(i)}^{*}\frac{T^{,m}}{T}\nonumber \\
 &  & -\frac{1}{3}\sum_{i}\int f_{(i)}^{(0)}\frac{1}{z_{(i)}}\left(\gamma_{k_{(i)}}-G\left(z_{(i)}\right)\right)K_{(i)}^{n}K_{n(i)}B_{(i)}d^{3}K_{(i)}^{*}V^{m}\\
 &  & -\frac{1}{3}\sum_{i}\int f_{(i)}^{(0)}\frac{1}{z_{(i)}}\left(\gamma_{k_{(i)}}-G\left(z_{(i)}\right)\right)K_{(i)}^{n}K_{n(i)}D_{(i)}d^{3}K_{(i)}^{*}\frac{1}{nh_{E}}\left(\frac{\tilde{\rho}_{(j)}}{z_{(i)}}\left(\nabla\mu_{(i)}\right)_{T}-\frac{n_{(j)}\tilde{\rho}_{(i)}}{n_{(i)}z_{(j)}}\left(\nabla\mu_{(j)}\right)_{T}\right),\nonumber \end{eqnarray}
which can also be written as\begin{equation}
\frac{q_{tot}^{,m}}{k_{B}T}=-L_{qq}^{*}\left[\frac{T^{,m}}{T}\right]-L_{qV}^{*}V^{m}-L_{q\mu}^{*}\frac{1}{nh_{E}}\left(\frac{\tilde{\rho}_{(j)}}{z_{(i)}}\left(\nabla\mu_{(i)}\right)_{T}-\frac{n_{(j)}\tilde{\rho}_{(i)}}{n_{(i)}z_{(j)}}\left(\nabla\mu_{(j)}\right)_{T}\right).\label{eq:58}\end{equation}
Similarly the volume flux can be written as,\begin{eqnarray}
J_{V}^{m} & = & -\frac{1}{3}\sum_{i}\int f_{(i)}^{(0)}\left(\gamma_{k_{(i)}}-G\left(z_{(i)}\right)\right)K_{(i)}^{n}K_{n(i)}\left(A_{(i)}+D_{(i)}\frac{n_{(j)}}{\tilde{\rho}}\left(m_{(j)}G\left(z_{(j)}\right)-m_{(i)}G\left(z_{(i)}\right)\right)\right)d^{3}K_{(i)}^{*}\frac{T^{,m}}{T}\nonumber \\
 &  & -\frac{1}{3}\sum_{i}\int f_{(i)}^{(0)}\left(\gamma_{k_{(i)}}-G\left(z_{(i)}\right)\right)K_{(i)}^{n}K_{n(i)}B_{(i)}d^{3}K_{(i)}^{*}V^{m}\\
 &  & -\frac{1}{3}\sum_{i}\int f_{(i)}^{(0)}\left(\gamma_{k_{(i)}}-G\left(z_{(i)}\right)\right)K_{(i)}^{n}K_{n(i)}D_{(i)}d^{3}K_{(i)}^{*}\frac{1}{nh_{E}}\left(\frac{\tilde{\rho}_{(j)}}{z_{(i)}}\left(\nabla\mu_{(i)}\right)_{T}-\frac{n_{(j)}\tilde{\rho}_{(i)}}{n_{(i)}z_{(j)}}\left(\nabla\mu_{(j)}\right)_{T}\right),\nonumber \end{eqnarray}
or\begin{equation}
J_{V}^{m}=-L_{Vq}^{*}\frac{T^{,m}}{T}-L_{VV}^{*}V^{m}-L_{V\mu}^{*}\frac{1}{nh_{E}}\left(\frac{\tilde{\rho}_{(j)}}{z_{(i)}}\left(\nabla\mu_{(i)}\right)_{T}-\frac{n_{(j)}\tilde{\rho}_{(i)}}{n_{(i)}z_{(j)}}\left(\nabla\mu_{(j)}\right)_{T}\right).\label{eq:59}\end{equation}
And lastly, the mass flux reads\begin{eqnarray}
\frac{J_{(i)}^{m}}{m_{(i)}} & = & -\frac{1}{3}\int\left(f_{(i)}^{(0)}K_{(i)}^{n}K_{n(i)}A_{(i)}+f_{(i)}^{(0)}K_{(i)}^{n}K_{n(i)}D_{(i)}\frac{n_{(j)}}{\tilde{\rho}}\left(m_{(j)}G\left(z_{(j)}\right)-m_{(i)}G\left(z_{(i)}\right)\right)\right)d^{3}K_{(i)}^{*}\frac{T^{,m}}{T}\nonumber \\
 &  & -\frac{1}{3}\int f_{(i)}^{(0)}K_{(i)}^{n}K_{n(i)}B_{(i)}d^{3}K_{(i)}^{*}V^{m}\nonumber \\
 &  & -\frac{1}{3}\int f_{(i)}^{(0)}K_{(i)}^{n}K_{n(i)}D_{(i)}d^{3}K_{(i)}^{*}\frac{1}{nh_{E}}\left(\frac{\tilde{\rho}_{(j)}}{z_{(i)}}\left(\nabla\mu_{(i)}\right)_{T}-\frac{n_{(j)}\tilde{\rho}_{(i)}}{n_{(i)}z_{(j)}}\left(\nabla\mu_{(j)}\right)_{T}\right),\end{eqnarray}
or\begin{equation}
\frac{J_{(i)}^{m}}{m_{(i)}}=-L_{dq}^{*}\frac{T^{,m}}{T}-L_{dV}^{*}V^{m}-L_{d\mu}^{*}\frac{1}{nh_{E}}\left(\frac{\tilde{\rho}_{(j)}}{z_{(i)}}\left(\nabla\mu_{(i)}\right)_{T}-\frac{n_{(j)}\tilde{\rho}_{(i)}}{n_{(i)}z_{(j)}}\left(\nabla\mu_{(j)}\right)_{T}\right).\label{eq:60}\end{equation}
In Eqs. (\ref{eq:58}), (\ref{eq:59}) and (\ref{eq:60}), when we
examine the integral expressions corresponding to those coefficients
subject to exhibit the appropriate symmetry which arises from the
same transformations used in the proof of the H theorem as well as
the results obtained in the previous section one immediately finds
that such symmetry does not hold. Therefore it turns out that,\begin{equation}
L_{dq}^{*}\neq L_{q\mu}^{*},\end{equation}
and the same for the others cross-coefficients. Thus, these calculations
clearly exhibit the fact that also in the relativistic case, the chemical
potentials of the species do not provide an adequate representation
in which the ORR are valid.

\section{Discussion}

In this paper we have shown that the introduction of the concept of
thermal velocity is equally useful to deal with transport properties
of diluted mixtures. In fact, the expression we obtained for the total
heat flux $J_{tot}^{m}$ is consistent with its expression in the
phenomenological theory as well as in the non-relativistic case. Secondly
we insist that the new result exhibits the existence of two representations
in which the ORR are valid. In the one discussed in section 4, where
the forces are those that have been used by other authors Refs. \cite{Groot-Leewen-Weert,CERCIGNANI}
for the simple component gas, is characterized by the fact that {}``Fourier's
like equation'' has to be modified by the presence of a pressure
gradient. 

In the second case as discussed in section 5, we propose the new idea
of the volume flux, which may be introduced without modifying the
classical Fourier equation, and also gives rise to the canonical form
for the Dufour and Soret effects related with $L_{qd}$ and $L_{dq}$.
This representation is new and provides two new cross effects that
are only present in the relativistic case namely, $L_{qV}$ and $L_{dV}$.
Indeed, one can immediately see from Eq. (\ref{eq:49}) that this
contribution vanishes in the non-relativistic limit.

Notice that the volume flow as introduced in Eq. (\ref{eq:53}) may
be regarded as a multiple of the heat flux Eq. (\ref{eq:53.1}) in
the single-fluid limit. As shown in the appendix, the constitutive
equation for the heat flux and for the volume flux in this limit coincide.
Thus, what in the binary mixture is a cross effect turns into a direct
effect with a Fourier type constitutive equation in the single-fluid
limit. 

The apparently new variable associated with the volume transport has
a peculiar thermodynamical meaning. This volume flux with its conjugated
force are indeed related with the thermodynamic description of the
system and when taken into account, clarify the nature of the transport
phenomena in a relativistic mixture. This coupling of the volume flux
with a pressure gradient is indeed confirmed when calculating the
entropy production of the mixture, which constitutes work in progress
and will be published elsewhere. 
\begin{acknowledgments}
The authors wish to thank Alfredo Sandoval-Villalbazo for his helpful
comments and Universidad Iberoamericana Ciudad de Mexico for hosting
part of this work. One of us, V. M. acknowledges CONACyT for financial
support under scholarship number 203111.
\end{acknowledgments}

\section*{Appendix}

In this appendix we will take the single-fluid limit from equations
for the volume flux and the heat flux. Taking $m_{(i)}=m_{(j)}=m$,
$n_{(i)}=n_{(j)}=n$, from Eqs. (\ref{eq:15}) and (\ref{eq: vol flow})
we have for the heat flux\begin{equation}
\frac{q_{tot}^{m}}{k_{B}T}=\frac{mc^{2}}{k_{B}T}\int\gamma_{k}K^{m}fd^{3}K^{*},\label{eq:A1}\end{equation}
and for the volume flux\begin{equation}
J_{V}^{m}=\int\gamma_{k}K^{m}fd^{3}K^{*},\label{eq:A2}\end{equation}
thus\begin{equation}
\frac{q_{tot}^{m}}{k_{B}T}=\frac{1}{z}J_{V}^{m}.\label{eq:A3}\end{equation}
It remains to verify that the transport coefficients satisfy the same
relations namely, from Eqs. (\ref{eq:53}) and (\ref{eq:53.1}) with
the fact that $d^{m}=0$, recalling that $J_{(i)}^{m}=J_{(j)}^{m}=0$
we get,\begin{equation}
J_{V}^{m}=-L_{Vq}\frac{T^{,m}}{T}-L_{VV}\left[\frac{nm}{\tilde{\rho}}\frac{p^{,m}}{p}\right]\label{eq:A4}\end{equation}
and\begin{equation}
\frac{q_{tot}^{m}}{k_{B}T}=-L_{qq}\frac{T^{,m}}{T}-L_{qV}\left[\frac{nm}{\tilde{\rho}}\frac{p^{,m}}{p}\right],\label{eq:A5}\end{equation}
where \begin{eqnarray}
L_{Vq} & = & -\frac{1}{3}\int f^{(0)}\left(\gamma_{k}-G\left(z\right)\right)K^{n}K_{n}Ad^{3}K^{*}\label{eq:A6}\\
L_{VV} & = & -\frac{1}{3}\int f^{(0)}\left(\gamma_{k}-G\left(z\right)\right)K^{n}K_{n}Bd^{3}K^{*}\label{eq:A7}\\
L_{qq} & = & -\frac{1}{3}\int f^{(0)}\frac{1}{z}\left(\gamma_{k}-G\left(z\right)\right)K^{n}K_{n}Ad^{3}K^{*}\label{eq:A8}\\
L_{qV} & = & -\frac{1}{3}\int f^{(0)}\frac{1}{z}\left(\gamma_{k}-G\left(z\right)\right)K^{n}K_{n}Bd^{3}K^{*}.\label{eq:A9}\end{eqnarray}
Where again we can immediately see that \begin{equation}
\frac{q_{tot}^{m}}{k_{B}T}=\frac{1}{z}J_{V}^{m}.\label{eq:A10}\end{equation}
Then, in the single-fluid limit, the volume flux turns out to be a
multiple of the heat flux.

\end{document}